# Electron and hole doping of monolayer WSe$_2$ induced by twisted ferroelectric hexagonal boron nitride

J. Frauni é[1], R. Jamil[1], R. Kantelberg[1], S. Roux[1], L. Petit[1], E. Lepleux[2], L. Pacheco[2], K. Watanabe[3], T. Taniguchi[4], V. Jacques[5], L. Lombez[1], M.M. Glazov[6], B. Lassagne[1], X. Marie[1], C. Robert[1†]

[1]Université de Toulouse, INSA-CNRS-UPS, LPCNO, 135 Av. Rangueil, 31077 Toulouse, France
[2]CSI Instruments, 17 Av des Andes, 91940 Les Ulis, France
[3]International Center for Materials Nanoarchitectonics, National Institute for Materials Science, 1-1 Namiki, Tsukuba 305-00044, Japan
[4]Research Center for Functional Materials, National Institute for Materials Science, 1-1 Namiki, Tsukuba 305-00044, Japan
[5]Laboratoire Charles Coulomb, Université de Montpellier and CNRS, 34095 Montpellier, France
[6]Ioffe Institute, 26 Polytechnicheskaya, 194021 Saint Petersburg, Russia

*For the past few years, 2D ferroelectric materials have attracted strong interest for their potential in future nanoelectronics devices. The recent discovery of 2D ferroelectricity in twisted layers of insulating hexagonal boron nitride, one of the most used 2D materials, has opened the route to its integration into complex van der Waals heterostructures combining hybrid properties. Here we show that opposite polarizations in ferroelectric domains of a folded hBN layer can imprint local n and p doping in a semiconducting transition metal dichalcogenide WSe$_2$ monolayer. We demonstrate that WSe$_2$ can be used as an optical probe of ferroelectricity in hBN and show that the doping density and type can be controlled with the position of the semiconductor with respect to the ferroelectric interface. Our results establish the ferroelectric hBN/WSe$_2$ van der Waals stacking as a promising optoelectronic structure.*

The spectacular interest for 2D materials for the past decade has been mainly triggered by the large platform they opened for the design of new structures with combined properties. Indeed, the class of 2D materials now extends to various 2D semiconductors, insulators, metals, magnets, superconductors, topological insulators, etc. Stacking of these 2D materials into so-called van der Waals (vdW) heterostructures has opened the route toward the design of complex hybrid structures combining 2D materials with radically different properties [1]. Beyond combining the individual properties of both materials in the same structure they also enable a mutual control of these properties through proximity effects at the interface. Among the various 2D hybrid structures that have been reported for the past years, monolayers (ML) of semiconducting transition metal dichalcogenide (TMD) such as MoS$_2$ or WSe$_2$ often play an important role due to their peculiar optical and spin properties. For instance, these materials have been coupled to 2D magnets [2,3] to study magnetic proximity effects or with graphene to induce spin-orbit coupling [4].

Recently, the class of 2D materials has extended with the discovery of 2D ferroelectricity [5]. The integration of 2D ferroelectrics (FE) with 2D semiconductors (SC) would be interesting for several perspectives. First, it can be used for the design of electronic devices such as ferroelectric field effect transistors (FeFET) [6]. Second, it can open new strategies to implement optoelectronic properties. For instance, electric fields induced by the 2D-FE can be used to tune locally the optical properties and the doping density of the 2D-SC. It can also pave the way towards the optical control of FE properties to adress FE-based memories. While many studies have reported the integration of TMD materials with standard bulk (3D) FE materials such as $BiFeO_3$ for the purpose of electronic devices [7–9], the literature on the optical properties of 3D-FE/TMD heterostructures is much scarcer. The reason is that these structures suffer from one main drawback: the 3D nature of the FE materials degrades the optical quality of the overlying TMD layer due to the presence of dangling bonds. A fully 2D-FE/TMD hybrid vdW heterostructure is thus highly demanded. While a few 2D materials (SnTe, $CuInP_2S_6$ …) exhibit intrinsic ferroelectricity due to their distorted crystal structure [10,11], their integration with TMD layers has been mostly limited to the demonstration of FeFET prototypes [12,13]. Very few studies have dealt with the optical properties of 2D-FE/TMD heterostructures. Mao *et al.* demonstrated hysteretic optical properties in a $MoSe_2$/$CuInP_2S_6$ gated device [14] while Li *et al.* recently measured the optical properties of a $MoS_2$/$CuInP_2S_6$ stacking but observed a strong quenching of the luminescence of $MoS_2$ due to the semiconducting nature of $CuInP_2S_6$ [15]. In summary, finding a 2D-FE material that can tune but not degrade the optical properties of the TMD layer would thus constitute a major breakthrough in the field.

Hexagonal boron nitride (hBN) is a 2D insulator that has been extensively used to encapsulate TMD MLs and strongly enhance their optical properties [16,17]. In its bulk form, hBN is not ferroelectric but recently, it has been demonstrated that ferroelectricity can occur at the interface between marginally twisted hBN layers close to a parallel alignment [18–21] opening the field of sliding ferroelectricity [22–24]. Promising theoretical works have predicted that the electric potential of twisted hBN can be imprinted in a TMD ML [25] resulting in large shifts of the band edges in the SC. Very recently, Kim *et al.* observed a change in the exciton diffusion in a $MoSe_2$ ML attributed to the proximity of a twisted hBN interface [26]. In this letter, we demonstrate experimentally that the presence of a FE-hBN interface with out-of-plane up and down polarized domains locally imprints electron and hole doped regions in the plane of a $WSe_2$ ML. In addition, we calculate the electric stray field induced by the FE interface and discuss its role on the doping mechanism and possible dissociation of excitons in the $WSe_2$ ML. Our results establish a new strategy to locally dope TMD layers, change their optical properties without using external metallic gates and more generally highlight the potential of FE-hBN/TMD heterostructures for the design of original optoelectronics properties.

Out-of-plane electric polarization can be created when two layers of hBN are artificially aligned in a parallel configuration with a twist angle close to 0°. Such a stacking generally results in reconstructed triangular domains with stable AB and BA arrangement that exhibit perpendicular dipoles with up and down electric polarization (see Figure 1a). In the previous studies, two main techniques have been used to fabricate these stacking. Yasuda *et al.* used the "tear-and-stack" technique to pick up one half of a hBN monolayer flake and transfer it on top of the remaining half [20]. Vizner Stern *et al.* used a very similar technique but on a slightly thicker flake (1-5nm) showing that the observation of ferroelectricity is not limited to twisted bilayers [19]. Finally, Woods *et al.* stacked two hBN flakes that were adjacently exfoliated on a $SiO_2$/Si substrate assuming that they were originated from the same growth domain and are thus nearly perfectly aligned [21]. Intentional or unintentional folding of 2D layers have been successfully used to create moiré pattern [27–29]. In our work, we show that we can obtain FE-hBN in a simple way by selecting thin hBN flakes that are accidentally folded during the exfoliation process. Figure 1c shows an optical image of a thin hBN flake (~13 nm) that was exfoliated using the scotch tape technique on a $SiO_2$(80nm)/Si surface (see details in Supplemental Material [30]). We clearly see a dark blue region corresponding to the thinnest part of the flake and a light blue region corresponding to a thicker part. The light blue region corresponds to the same thin flake that has been folded probably during the exfoliation process [34]. Interestingly, we clearly see that the edges of the folded part make a perfect 60° angle that are likely representing either zigzag or armchair crystallographic axes of hBN [35]. In addition, we note that the angles between the folding axis and the edges of the flake are also 60° so that these three directions correspond to the same configuration (they are either all zigzag or all armchair). We show in the Supplemental Material [30] that folding along a zigzag (armchair) axis necessarily correspond to an antiparallel (parallel) alignment of the two hBN layers at the interface. The folded part is then characterized using atomic force microscopy (AFM) and Kelvin probe force microscopy (KFM). The AFM image (Figure 1d) shows that the top surface of the flake is atomically flat. The KFM image (Figure 1e) shows triangular regions with two distinct surface potentials that are characteristic of a parallel alignment and ferroelectric domains with AB and BA atomic arrangement at the interface between the two folded parts of the flakes [19–21]. This demonstrates that the folding axis in this flake is along an armchair direction. The potential difference between bright and dark domains is $150 \pm 20$ mV in agreement with previous studies [19–21]. The typical size of domains varies from 100 nm to 1 µm on this flake. Other examples of folded flakes exhibiting small triangular domains are shown in the Supplemental Material [30]. This first result demonstrates that we can easily obtain FE domains using a single step mechanical exfoliation technique by selecting self-folded flakes. Beyond its simplicity as compared to the stacking methods previously reported in the literature [19–21], this technique can potentially create very large domains exceeding the micrometer size. Figure 2b shows the KFM image of a another folded hBN flake with similar surface potential contrast but with larger area. Other examples are shown in Figure 4b and in the Supplemental Material [30]. In the following, we will use such hBN flakes with large domains to optically probe in-plane doping modulation in TMD ML in close proximity.

Figure 2a shows a sketch of the studied structure. A $WSe_2$ ML is transferred on top of the folded hBN flake with large FE domains and capped with a thin (not folded) hBN flake (see the fabrication details in the Supplemental Material [30]). The distance between the $WSe_2$ ML and the FE interface is simply given by the thickness of the folded part (10.5 nm for this sample). We checked that the transfer of the $WSe_2$ ML and the top hBN flake does not significantly affect the size of the FE domains by performing KFM measurements at each step of the fabrication (see Supplemental Material [30]). We then perform micro-photoluminescence (PL) measurements on the $WSe_2$ ML. Importantly, the diameter of the optical spot is limited by diffraction (~500 nm) but is smaller than the size of the FE domains so we can measure the PL of $WSe_2$ on top of a *bright* or a *dark FE* domain separately (here *bright* and *dark FE* domains refer to the KFM image of Figure 2b that we attribute to BA and AB stacking). Figure 2c presents typical PL spectra taken at three positions of the sample (on a *bright FE* domain, a *dark FE* domain and on the unfolded part of the hBN flake showing *no FE* domain). Measurements are shown at 55 K and for two excitation powers of 50 nW and 50 µW. We recall in the Supplemental Material [30] the configurations of the different excitonic complexes that are observed in these spectra. We use the energy splitting of the peaks with respect to the bright neutral exciton and the power dependence of their intensity to identify the nature of the excitonic transitions in accordance with literature [36,37]. The key observations in Figure 2c are the following:

- The PL spectrum on the *no FE* area is characteristic of an intrinsically slightly n-doped $WSe_2$ ML [36,37]. At low excitation power (50 nW), we observe mainly the peak labelled $X^0$ that corresponds to the bright neutral exciton. At an energy 18 meV below $X^0$ we observe the neutral biexciton ($XX^0$) that can only be seen at high excitation power (50 µW). The two negative bright trions (triplet $X_T^-$ and singlet $X_S^-$) are observed 29 meV and 36 meV below $X^0$. Finally, the negatively charged biexciton ($XX^-$) is observed 51 meV below $X^0$ only at high excitation power. These negatively charged exciton complexes ($X_T^-$, $X_S^-$ and $XX^-$) can only be seen if the WSe$_2$ ML is n-doped.

- The PL spectrum on the *bright FE* domains shows the same peaks but with intensities that are characteristic of a larger n-doping. Indeed, the intensity of the neutral transitions ($X^0$ and $XX^0$) is reduced as compared to the *no FE* spectra while the intensity associated with negatively charged excitonic complexes ($X_T^-$, $X_S^-$ and $XX^-$) is enhanced.

- The PL spectrum on the *dark FE* domain is radically different. The negatively charged excitonic complexes are absent. On the opposite, we observe a peak 21 meV below $X^0$ at both high and low excitation powers. This peak is attributed to the positive bright trion ($X^+$) indicating a p-doping of the WSe$_2$ ML.

The results of Figure 2 clearly show that the FE domains in the bottom hBN flake control the nature of the doping in the WSe$_2$ ML (n-doping in *bright FE* domain versus p-doping in *dark FE* domain). In Figure 3, we show that the WSe$_2$ ML can be used to optically image the FE domains in hBN. By moving the sample with a piezo-driven scanner, we record the PL spectrum at each point of the WSe$_2$ ML. Measurements are performed at 55 K for an excitation power of 20 µW. For each pixel, we plot the ratio between the intensity of the negatively charged biexciton ($I_{XX^-}$) that is a signature of n-doping and the intensity of the positive trion ($I_{X^+}$) that is a signature of p-doping. We choose these two transitions because they are spectrally well separated. More details about the data analysis can be found in the Supplemental Material [30]. We clearly see that the PL ratio $I_{XX^-}/I_{X^+}$ is large (small) in the *bright (dark) FE* domains shown in the KFM image of Figure 2b. These results confirm that FE domains in hBN imprint n/p domains in the WSe$_2$ ML.

We now show that the type (p/n) and the density of doping can be controlled with the position of the WSe$_2$ ML with respect to the FE-hBN interface. Figure 4a shows the sketch of a second heterostructure where a folded hBN flake with FE domains is transferred onto of a WSe$_2$ ML while the bottom hBN flake is not folded. The distance between the FE interface and the WSe$_2$ ML is larger than in the structure presented in Figure 2 (15.8 nm). Figure 4b shows the KFM image of the final structure. Although the FE domains are significantly affected by the transfer process, we can identify two large FE domains with opposite polarization in the top hBN flake. Figure 4c presents the PL spectra of the WSe$_2$ ML in both regions. The dark domain in the KFM image now corresponds to a n-doped WSe$_2$. This is opposite to the results of Figure 2 but perfectly consistent with the direction of the electric field in the WSe$_2$ ML; i.e. in Figure 2 the FE-hBN is below the WSe$_2$ while it is above in Figure 4. The PL spectrum taken on the bright FE domain is typical of a nearly neutral WSe$_2$ ML where charged excitonic complexes are absent and only neutral transitions (bright exciton $X^0$, biexciton $XX^0$ and dark exciton $X^D$) are visible. Given that the WSe$_2$ ML is intrinsically slightly n-doped, we conclude that the effect of the bright FE domain is to slightly compensate the doping with holes. The results of Figure 4 demonstrate that the 2D-SC layer can feel the influence of the FE-hBN at a distance as large as 15 nm but with a p/n modulation that is smaller than the modulation observed in the structure of Figure 2. This could be due to the larger distance between the FE-hBN interface and the

WSe$_2$ ML and shows that the doping density in WSe$_2$ can potentially be tuned with the thickness of the folded hBN flake. Nevertheless, further studies are required to properly separate the effect of the hBN thickness on the doping density from possible extrinsic parameters such as strain, contamination or defect density in the ML.

While our results clearly demonstrate a local modulation of the doping from n-type to p-type in the WSe$_2$ ML, they raise the question about the origin of the free charge carriers. Indeed, despite its FE interface, hBN remains an insulator and it is unlikely that charge transfer occurs between WSe$_2$ and hBN. Possible piezoelectric effects in our structure due to local strain seem to be unlikely owing to reduced strain in the WSe$_2$ and weak piezoelectricity in multi-layers of hBN. A possibility is that the electric stray field induced by the FE interface favors the release of charges that are trapped in the WSe$_2$ ML or in close proximity, e.g. at the interface with hBN. This scenario is consistent with the temperature dependence of the doping density in our FE-hBN/WSe$_2$ heterostructure (see Supplemental Material [30]). When the sample is cooled down to 5 K, the modulation of the doping between *bright* and *dark FE* domains occurs but is reduced as compared to the results of Figure 2 shown at 55 K. Both thermal energy and electric field may contribute to the release of trapped charges.

Another possibility is that n-doped and p-doped regions result from the combination of light excitation and in-plane electric field. Indeed, light excitation of the WSe$_2$ ML generates electron-hole pairs that bind to form neutral excitons with a significant binding energy (around 170 meV [38] for a WSe$_2$ ML encapsulated into hBN). These excitons can dissociate only under a large in-plane electric field. In FE-hBN domains, the direction of the electric field is out-of-plane. Nevertheless, the stray electric field that is present in the WSe$_2$ ML above or below a FE domain wall is necessarily in-plane [25]. Depending on the distance between the FE interface and the ML, the amplitude of this in-plane electric stray field may be enough to partially dissociate excitons that are created in the vicinity of the domain wall so that free electrons and free holes accumulate on each side. These free charges can then bind to neutral excitons that are photogenerated in the domains to form the charged excitonic complexes observed in Figure 2c and 4c.

In both scenarios, the role of the electric stray field is crucial. It is thus important to estimate its amplitude using a simple model based on Maxwell equations. We calculate the electric field induced by a FE domain wall caused by the two stackings AB and BA (see the sketch in Figure 5a). We assume that the FE interface is in the $(xy)$ plane and is surrounded by a homogeneous medium of hBN (not ferroelectric) with permittivity $\varepsilon$. For simplicity, we assume that the domain wall is infinitely small at $x = 0$ and do not take into account any in-plane polarization around the domain wall [39]. We consider two semi-infinite domains of opposite dipole moment density $\pm P_0$ parallel to the $z$ axis. This assumption is realistic as the size of domains in our samples is much larger than the distance between the FE interface and the WSe$_2$ ML. We can show (see Supplemental Material [30] for calculation details) that the electric field components are:

$$E_x = -\frac{4P_0}{\varepsilon}\frac{z}{x^2+z^2}, \; E_y = 0 \; , E_z = \frac{4P_0}{\varepsilon}\frac{x}{x^2+z^2}$$

Figure 5b presents both in-plane ($E_x$) and out-of-plane ($E_z$) components as a function of $x$ for a distance $z =$ 10 nm corresponding to the position of the WSe$_2$ ML in the sample presented in Figure 2. As expected, the electric field is strictly in-plane just above the domain wall ($x = 0$) whereas the out-of-plane component varies from positive to negative while crossing the domain wall. The maximum in-plane electric field is 30 mV.nm$^{-1}$. This is comparable to typical external electric field used in photocurrent measurements that evidenced exciton dissociation in WSe$_2$ [40] and not far from the theoretical value of 50 mV.nm$^{-1}$ corresponding to equal dissociation rate and intrinsic decay rate of the exciton in hBN/MoS$_2$/hBN [41]. Moreover, it was shown recently that 10% of excitons can spontaneously dissociate into electron-hole pairs due to interaction with defects [42]. The in-plane electric field present in our samples could thus assist the separation of electrons and holes into separate domains. Finally, the in-plane electric field may also favor dissociation of the excited states of excitons with smaller binding energy. In summary, our calculations prove that exciton dissociation at the domain wall must be considered as a possible mechanism for creating the charges. It is also important to note that both electric field components vanish for $x \to \infty$. This means that the electric stray field is almost zero in the WSe$_2$ ML located at the center of the domains. On the opposite, the results of Figure 3 show that large areas of WSe$_2$ are charged and not only the surrounding of the domain walls. This proves that whatever the mechanism that create these charges is, diffusion occurs on distances as long as the domain size. Moreover, it is important to note that in our samples the WSe$_2$ ML is not connected to any electrode so that accumulated charges in separate domains cannot be easily evacuated.

Finally, we present in Figure 5c, the amplitude of the maximum in-plane electric field (at $x = 0$) as a function of $z$. It clearly shows that reducing the distance between the WSe$_2$ ML and the FE interface, i.e. reducing the thickness of hBN down to a few layers, should strongly enhance the electric field in the range of a few 100's of mV.nm$^{-1}$. Such a field is difficult to obtain with external electrodes and it would be interesting for many perspectives where efficient dissociation of excitons is needed including photodetector, photovoltaics, photocatalysis and other applications.

In summary, we have shown that parallel alignment can occur when a hBN flake is folded along one of its armchair axes resulting in ferroelectric domains at the interface. We then demonstrated that ferroelectric hBN can imprint in plane p/n charge carriers domains in a WSe$_2$ ML. Such heterostructures could be used in the near future for several applied and fundamental perspectives. Original architectures of LED or photodetectors can be built by depositing electric contacts on both p and n regions. Besides exciton dissociation, the combination of p/n doping in the TMD and the in-plane electric field in the vicinity of a domain wall may result in the localization of one-dimensional excitons similar to the recent results of Thureja *et al.* [43] using external gates. Further nontrivial localization effects may also occur at the apexes of the triangular domains. Finally, our results should stimulate future studies on possible optical switching of the FE polarization in hBN using photogenerated excitons in a TMD structure as it was demonstrated in 3D-FE/MoS$_2$ [44].

## Acknowledgements

We thank T. Amand for fruitful discussions. This work was supported by Agence Nationale de la Recherche funding under the program ESR/EquipEx+ (grant number ANR-21-ESRE-0025), ANR ATOEMS and ANR IXTASE, through the grant NanoX n° ANR-17-EURE-0009 in the framework of the "Programme des Investissements d'Avenir" and by the Institute for Quantum Technologies in Occitanie through the project 2D-QSens. K.W. and T.T. acknowledge support from the Elemental Strategy Initiative conducted by the MEXT, Japan and the CREST (JPMJCR15F3), JST. X.M. also acknowledges the Institut Universitaire de France.

† Corresponding author: cerobert@insa-toulouse.fr

**Figures**

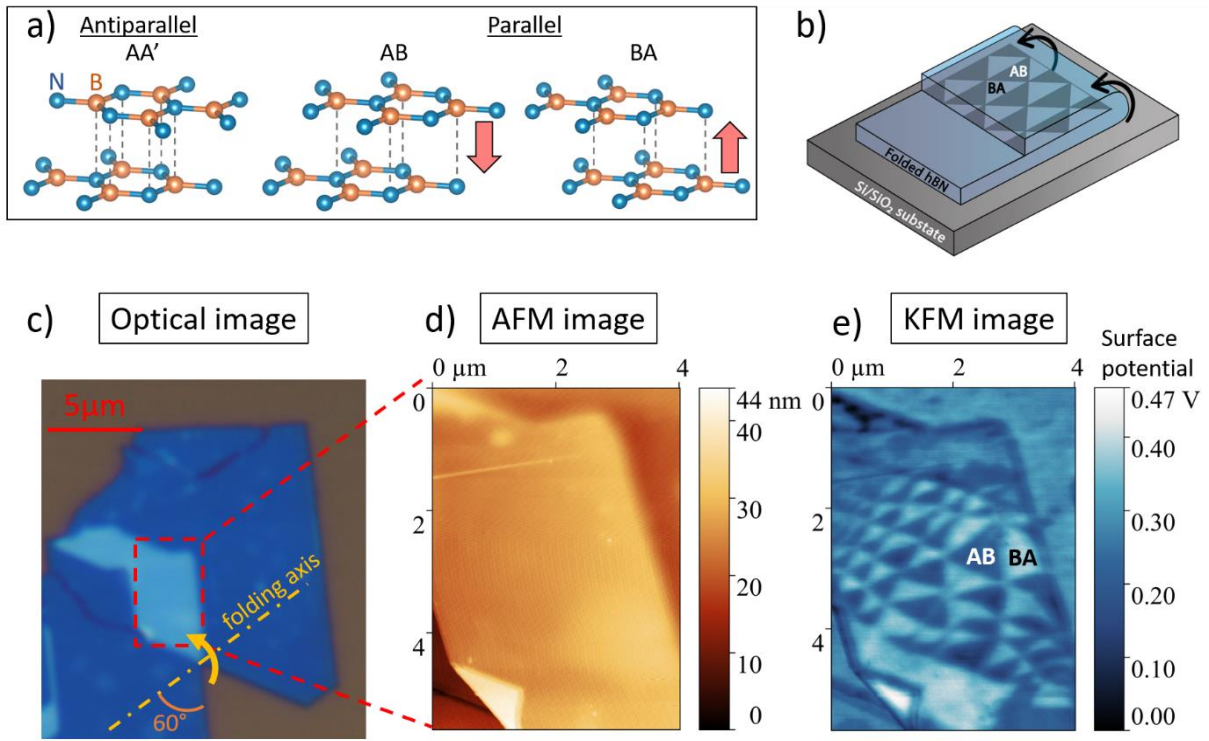

*Figure 1:a) Stable stacking arrangements for antiparallel and parallel alignment. The AA' stacking is the configuration in bulk hBN. For parallel alignment, only AB and BA stacking are stable and exhibit out-of-plane electric polarization marked by the red arrows. b) Sketch of a folded hBN flake. Triangular AB and BA ferroelectric domains are present at the folded interface. c,d,e) Optical, AFM and KFM images of a hBN flake folded along an armchair axis.*

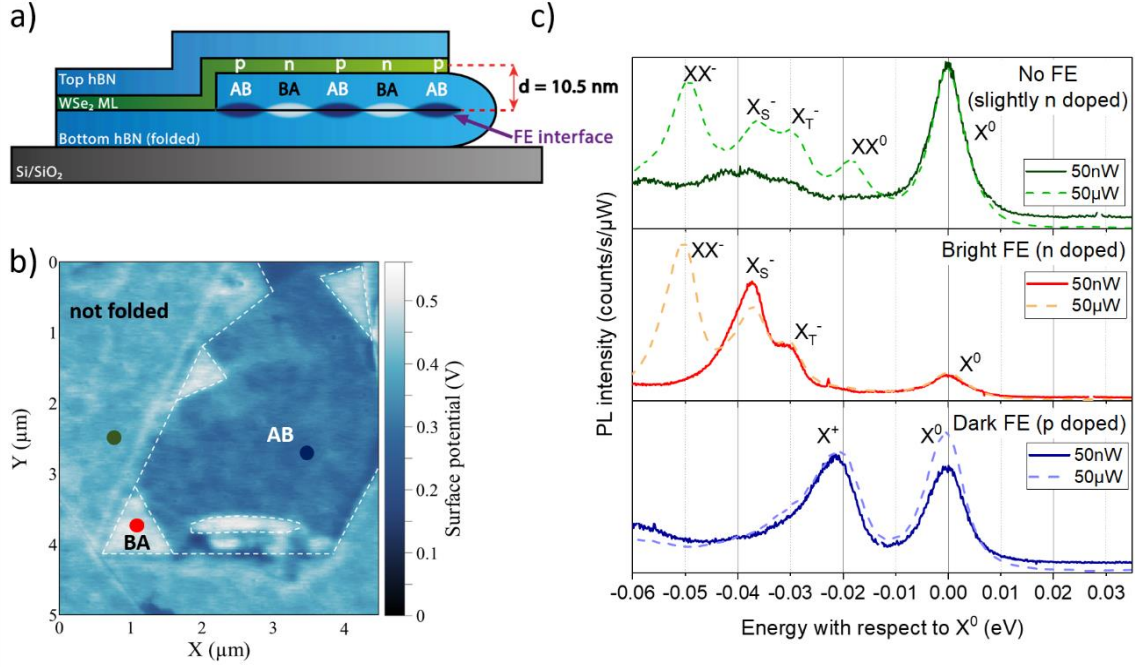

*Figure 2: Sketch of the FE-hBN/WSe$_2$ ML/hBN van der Waals heterostructure. b) KFM image of the folded bottom hBN flake. Bright (BA) and dark (AB) are clearly observed. White dashed lines show the borders of the ferroelectric domains. c) PL spectra at three typical positions of the sample marked by the colored points in b). Measurements are performed at T=55 K and for two extreme excitation powers of 50µW and 50 nW. For a given position, we divide the intensity by the integration time and the excitation power. Spectra are plotted as a function of the energy with respect to the bright neutral exciton $X^0$ peak to correct from energy shift of the whole spectra due to inhomogeneous strain.*

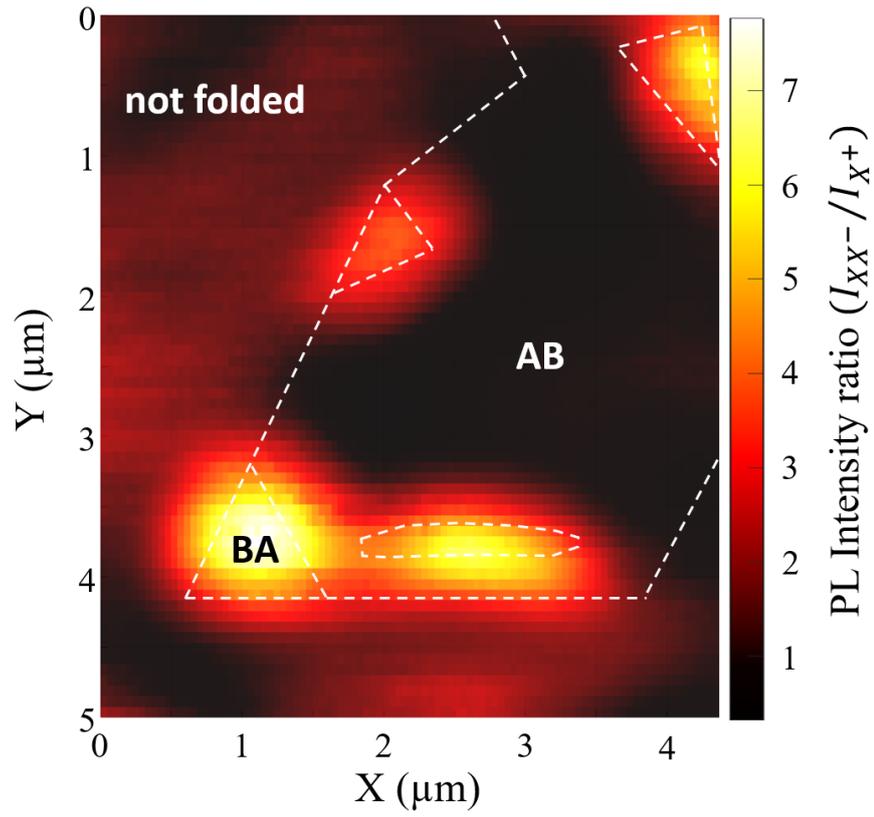

*Figure 3: Ratio of PL intensities of the negatively charged biexciton $I_{XX^-}$ and positive trion $I_{X^+}$ for the heterostructure presented in Figure 2. The borders of the ferroelectric domains observed in the KFM image of Figure 2b) are reproduced by white dashed lines.*

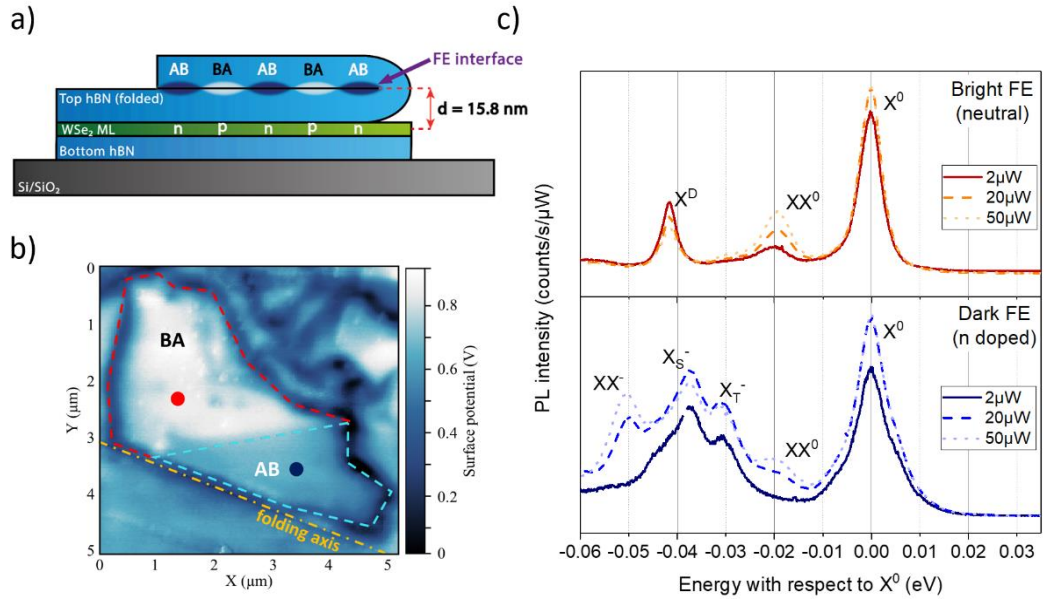

*Figure 4: Sketch of the hBN/WSe$_2$ ML/ FE-hBN van der Waals heterostructure. b) KFM image of the final structure. Bright (BA) and dark (AB) are highlighted with red and blue dashed lines. Optical and AFM images are presented in Supplementary Information) Typical PL spectra on both bright and dark ferroelectric domains marked by the colored points in b). Measurements are performed at T=55 K and for three excitation powers of 2µW, 20 µW and 50 µW. For a given position, we divide the intensity by the integration time and the excitation power. Spectra are plotted as a function of the energy with respect to the bright neutral exciton $X^0$ peak to correct from energy shift of the whole spectra due to inhomogeneous strain.*

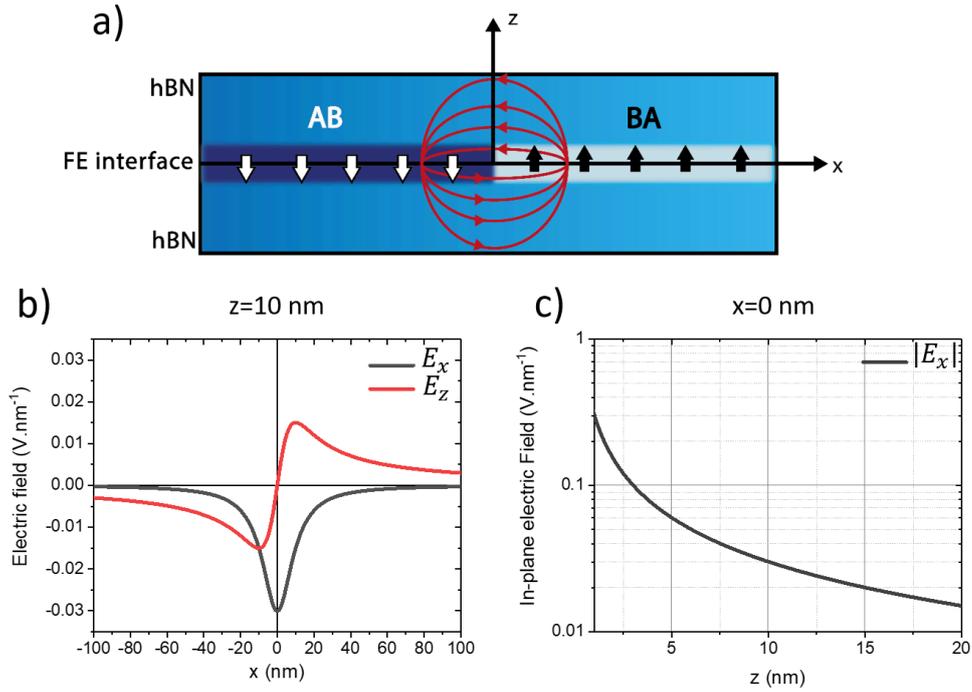

*Figure 5: a) Sketch of the structure used for the calculation of the electric stray field (red lines). We consider two semi-infinite planes with dipole moment density $+P_0$ (black arrows) and $-P_0$ (white arrows) separated by an infinitely small domain wall at x=0. We use $P_0 = 2$ pC.m$^{-1}$ [18]. We consider that the medium surrounding the FE interface is hBN with relative permittivity $\varepsilon_r = 3$. b) Calculated components of the electric field at a distance z=10 nm above the FE interface. c) Absolute value of the in-plane electric field as a function of z and for x=0 nm.*

# Supplemental Material for "Electron and hole doping of monolayer WSe$_2$ induced by twisted ferroelectric hexagonal boron nitride"

J. Fraunié[1], R. Jamil[1], R. Kantelberg[1], S. Roux[1], L. Petit[1], E. Lepleux[2], L. Pacheco[2], K. Watanabe[3], T. Taniguchi[4], V. Jacques[5], L. Lombez[1], M.M. Glazov[6], B. Lassagne[1], X. Marie[1], C. Robert[1]

[1]*Université de Toulouse, INSA-CNRS-UPS, LPCNO, 135 Av. Rangueil, 31077 Toulouse, France*
[2]*CSI Instruments, 17 Av des Andes, 91940 Les Ulis, France*
[3]*International Center for Materials Nanoarchitectonics, National Institute for Materials Science, 1-1 Namiki, Tsukuba 305-00044, Japan*
[4]*Research Center for Functional Materials, National Institute for Materials Science, 1-1 Namiki, Tsukuba 305-00044, Japan*
[5]*Laboratoire Charles Coulomb, Université de Montpellier and CNRS, 34095 Montpellier, France*
[6]*Ioffe Institute, 26 Polytechnicheskaya, 194021 Saint Petersburg, Russia*

## S1. Experimental methods

High quality hBN crystals [1] are exfoliated using the scotch tape technique on Si substrates covered by 83 nm of SiO$_2$. Thin folded flakes along a particular axis are identified by the contrast in an optical microscope and by measuring the angle between the folding axis and edges of the flake.

AFM and High-Definition KFM (HD-KFM) measurements are performed at room temperature using a Nano-Observer commercial microscope from CSI Instruments and Si probes coated with Pt. Topography and surface potential images are acquired simultaneously (single pass mode) using a highly sensitive bimodal approach where the electric feedback is tuned to the second eigenmode frequency of the cantilever.

The heterostructure shown in Figure 2 and Figure 3 is fabricated by transferring a WSe$_2$ ML and a thin top hBN flake (8 nm thick) on top of the folded FE-hBN flake by a dry stamping technique [2]. WSe$_2$ bulk crystal is purchased from HQ graphene. The heterostructure shown in Figure 4 is fabricated by first transferring a thick bottom hBN flake (~120 nm) followed by a WSe$_2$ ML using the dry stamping technique. The top folded FE-hBN is then released from its substrate and transferred on top of the WSe$_2$ ML using a polycarbonate (PC) assisted pick-up technique [3]. Both heterostructures are finally annealed on a hot plate at 150 °C during 1h30 to improve the interface quality.

Micro-photoluminescence experiments at low temperature are carried out in an ultra-stable confocal microscope. A continuous wave HeNe laser is coupled to a single mode fiber and the beam is focused on the sample using a high numerical aperture objective (NA=0.82). The luminescence is then coupled to a second single mode fiber and directed to a monochromator equipped with a Si-CCD camera. The size of both excitation and detection spots is limited by diffraction (~500 nm). The sample is moved with respect to the laser spot using a xyz piezo-driven scanner (for the mapping) and three x,y,z piezo-driven steppers (for coarse moving) from Attocube.

## S2. Determination of the folding axis

We show in this section that parallel alignment and thus ferroelectricity can occur in a folded hBN flake only if the folding axis is along an armchair crystallographic direction.

We first highlight in Figure S1a the angles between edges of the exfoliated flake presented in Figure 1c of the main text. In a honeycomb lattice, both armchair and zigzag directions have a periodicity of 60° and are rotated from each other by 30°. We can thus classify edges of the flakes into two categories marked by the yellow and pink dashed lines. We then present in Figure S1b and Figure S1c the stacking of two honeycomb lattices that are obtained by mirror symmetry along an armchair axis and a zigzag axis. We clearly see that the armchair mirror symmetry axis corresponds to a parallel alignment whereas the zigzag mirror symmetry axis corresponds to an antiparallel alignment.

Because only parallel alignment gives rise to ferroelectricity we conclude that observing AB and BA domains in KFM is due to a folding along an armchair axis.

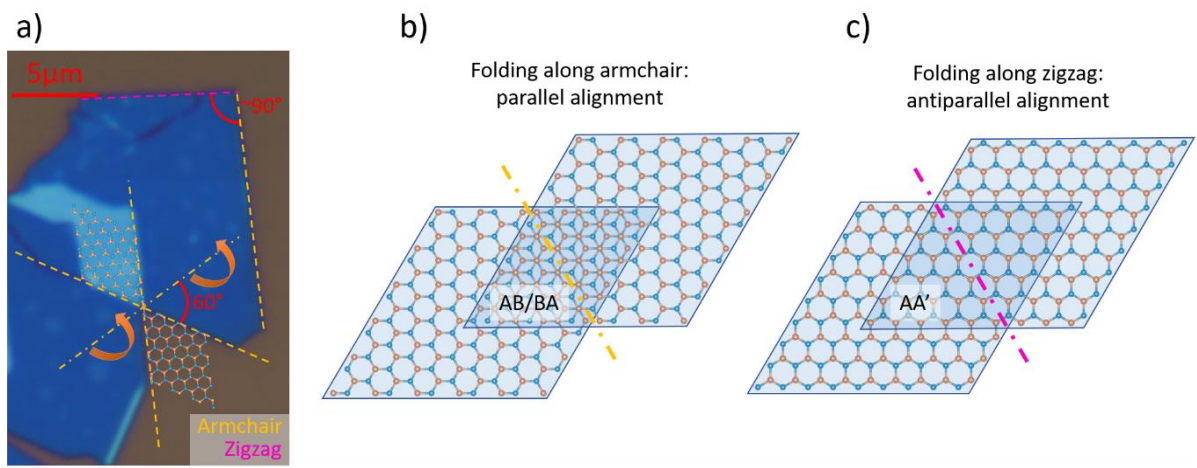

*Figure S1: a) Optical image of Figure 1c of the main text where we show armchair and zigzag edges. b) Folding along an armchair axis. c) Folding along a zigzag axis.*

## S3. Other examples of folded hBN flakes exhibiting ferroelectricity

We show in this section other examples of folded flakes with small and large ferroelectric domains. Note that our technique cannot control the size of the domains.

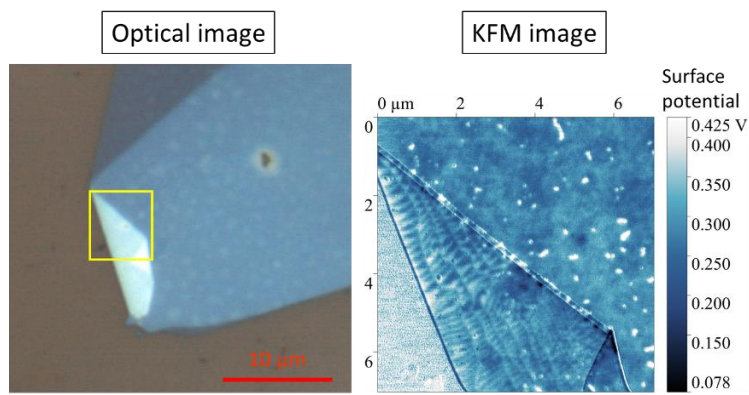

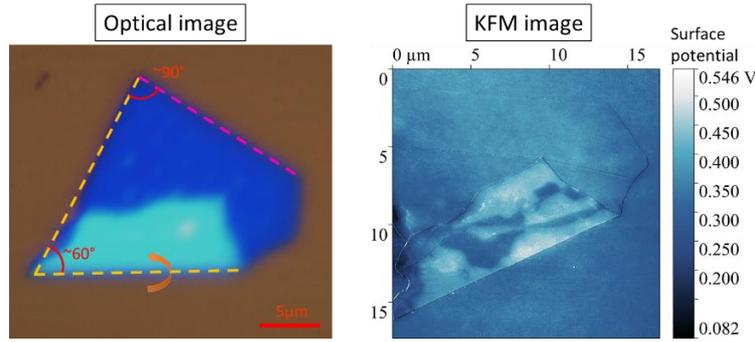

*Figure S2: Optical and KFM images of two folded hBN flakes with small triangular domains (top) and large domains (bottom)*

## S4. Optical and KFM images after WSe$_2$ and top hBN transfer

We show in Figure S3 the KFM images of the sample presented in Figure 2 and 3 of the main text before and after the transfer of the WSe$_2$ ML and the top hBN flake. While the deposition of WSe$_2$ and top hBN flakes results in blurred KFM imaging of the underlying ferroelectric domains, we clearly see that their shape is not significantly modified.

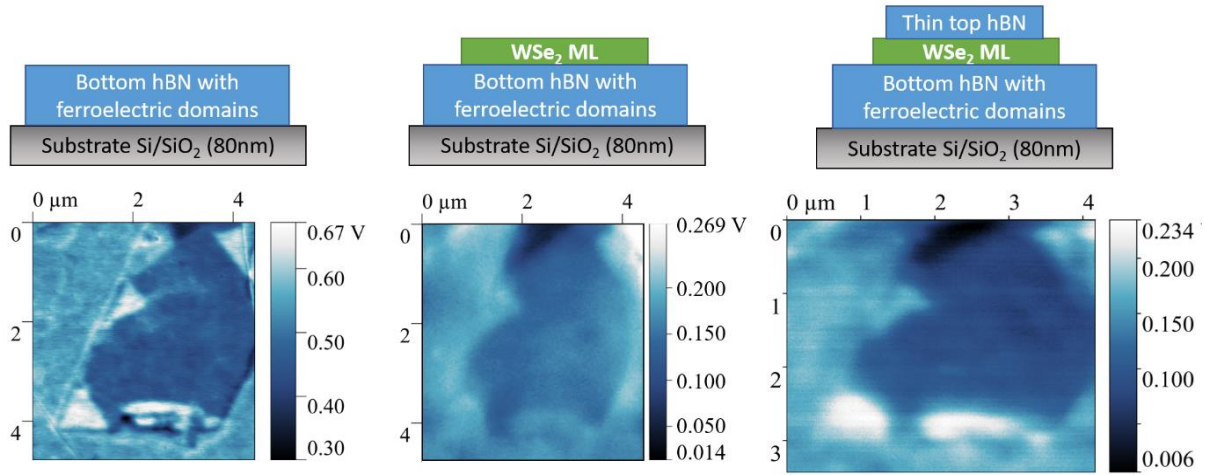

*Figure S3: KFM images of the structure of Figure 2 in the main text at each step of fabrication.*

## S5. Sketches of the excitonic complexes in WSe$_2$ monolayer

We recall in this section the configurations of the excitonic complexes in a WSe$_2$ ML mentioned in the main text. More details can be found in [4].

TMD MLs are direct band gap semiconductors with a band gap located at the K points of the hexagonal Brillouin zone. Due to broken inversion symmetry and strong spin-orbit coupling, both conduction bands and valence bands exhibit a large spin orbit splitting (a few 10's meV for conduction band and several 100's meV for the valence band). This peculiar band structure gives rise to various excitonic complexes (neutral or charged). Some of them are explained in Figure S4.

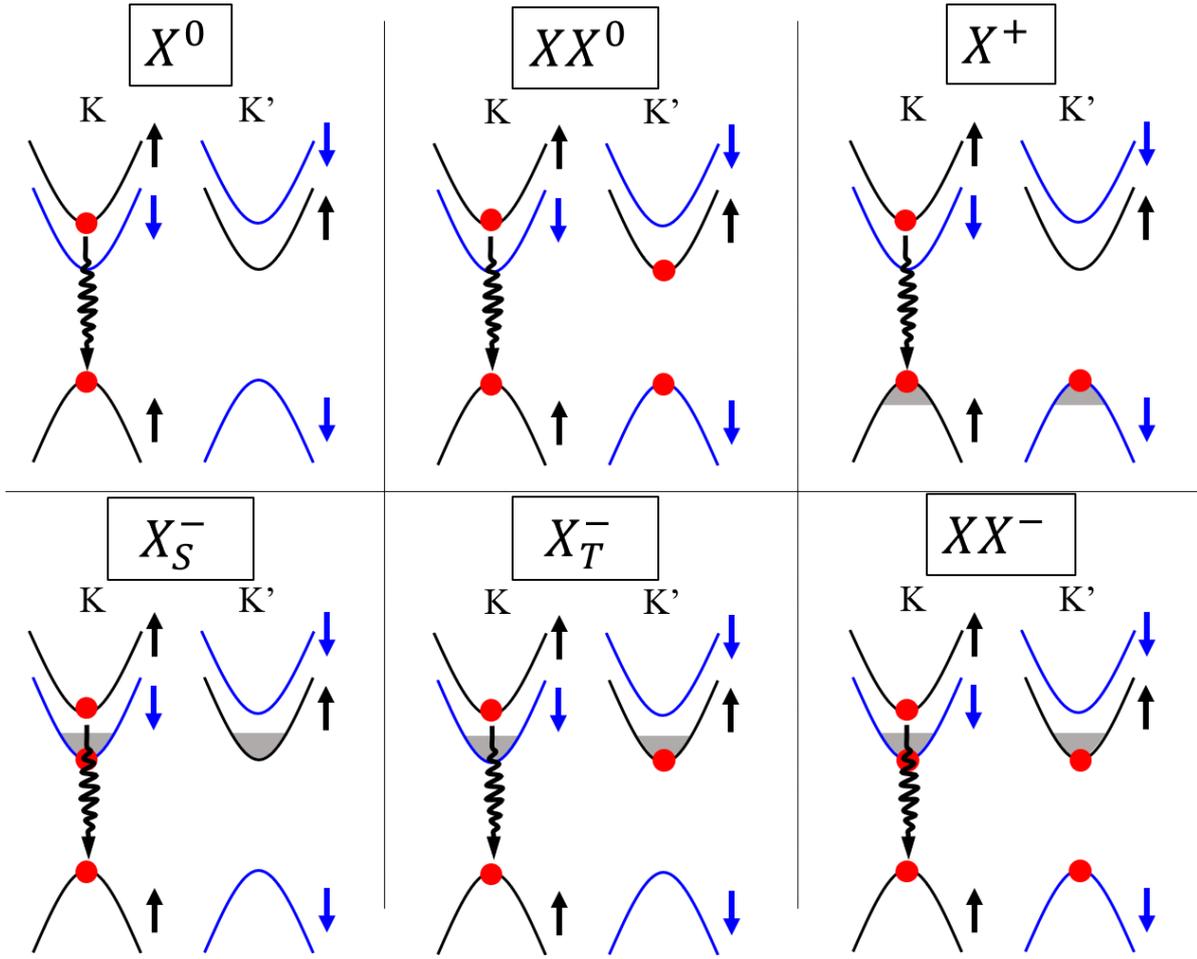

*Figure S4: Simplified band structure of WSe$_2$ ML. p-doping and n-doping are represented by shaded area in the first conduction band or first valence band. Emission of the photon is marked by the wavy black arrow.*

## S6. Protocol used for the analysis of photoluminescence mapping

We present in this section how we plot the Figure 3 of the main text. A PL spectrum is taken at each position of the scan. Because of inhomogeneities, PL spectra can be shifted in energy from point to point. Thus, for each spectrum we take the reference of energy at the peak of the bright neutral exciton $X^0$ transition. We then calculate the integrated PL intensity $I_{XX^-}$ for an energy window $51 \pm 4$ meV below $X^0$ that corresponds to the negatively charged exciton $XX^-$ in n-doped regions (red region in Figure S5). We also calculate for each spectrum the integrated PL intensity $I_{X^+}$ for an energy window $21 \pm 4$ meV below $X^0$ that corresponds to the positive trion $X^+$ in p-doped regions (blue region in Figure S5). Figure 3 of the main text plot the ratio $I_{XX^-}/I_{X^+}$.

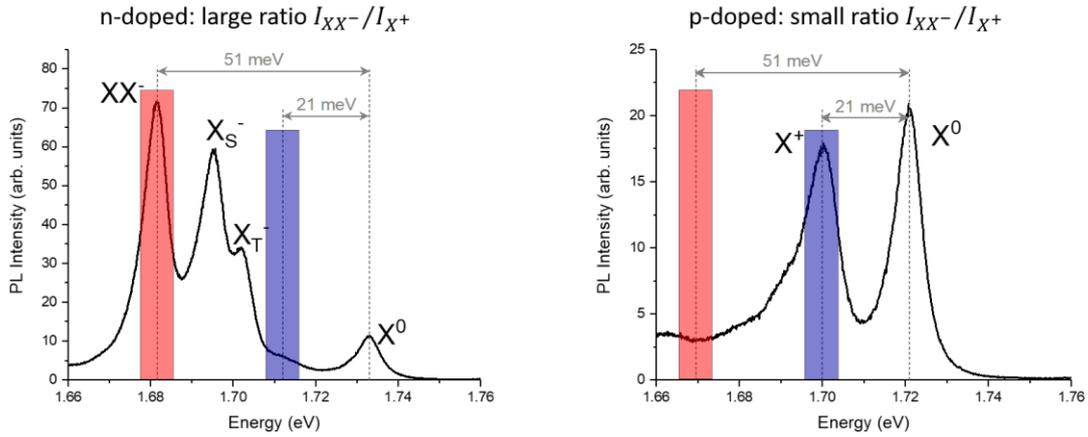

*Figure S5: Examples of two spectra in n-doped and p-doped regions. Red and blue rectangles show the energy ranges that are used to calculate the integrated PL intensities $I_{XX^-}$ and $I_{X^+}$.*

## S7. Temperature dependence of photoluminescence

PL spectra in the main text are presented at T=55 K. We present in Figure S6 the temperature dependence of the PL at two specific spots in the sample of Figure 2: on the dark ferroelectric domain (p-doped region) and on a bright domain (n-doped region). We plot for each temperature the PL intensity normalized by the peak intensity of the neutral exciton $X^0$ for the low excitation power of 50 nW. We clearly see that the positive trion $X^+$ appears only above 40 K with a maximum ratio $X^+/X^0$ around 55 K. At higher temperature, $X^+$ drops due to the activation of non-radiative channels as classically observed in TMD monolayers. Similar observations can be drawn for negative trions $X_S^-$ and $X_T^-$. Even if they can be observed at the lowest temperature of 10 K we clearly see that their intensity with respect to $X^0$ is maximum around 55 K and drops at higher temperature. These results show that the doping induced by the ferroelectric domains is thermally activated.

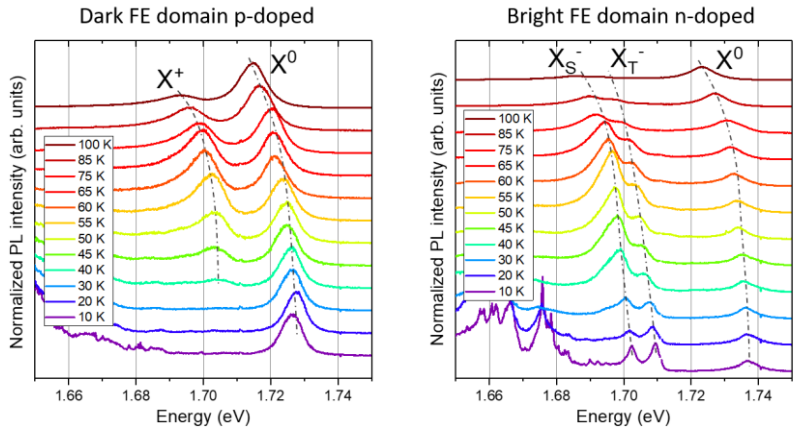

*Figure S6: Temperature dependence of PL on dark (p-doped) and bright (n-doped) FE domains. Each spectrum is normalized by the peak intensity of the $X^0$ transition. Excitation power is 50 nW. Dashed lines are guide to the eyes to highlight the energy of trions and neutral exciton.*

## S8. Details of the calculation of the electric stray field

We calculate the electric field induced by a ferroelectric domain wall caused by the interface between the two stackings of hBN, namely AB and BA. We assume that the hBN layers are stacked in the $(xy)$ plane and the domain wall is at the $x = 0$. We use the quasi-bulk model assuming that the system is infinite along the $z$-axis and treat the stacking as a two-dimensional plane with the two-dimensional dipole moment density $+P_0$ at $x > 0$ and $-P_0$ at $x < 0$:

$$\boldsymbol{P}(x,z) = P_0 \delta(z) \, \text{sign} x \, \hat{\boldsymbol{z}} \tag{1}$$

It follows from the Maxwell equations (written in CGS units) that:

$$\text{div} \, \boldsymbol{D} = 0 \Rightarrow \varepsilon \Delta \varphi(x,z) = 4\pi \, \text{div} \, \boldsymbol{P}(x,z) \tag{2}$$

Here $\varphi(x,z)$ is the electrostatic potential, $\boldsymbol{E} = -\boldsymbol{\nabla}\varphi$ and $\boldsymbol{D} = \boldsymbol{E} + 4\pi\boldsymbol{P}$, and $\varepsilon$ is the static dielectric constant.

Equation (2) can be conveniently solved by making a Fourier transform. We note that

$$\widetilde{\boldsymbol{P}}_{q_x, q_z} = \int_{-\infty}^{\infty}\int_{-\infty}^{\infty} dx dz \, \boldsymbol{P}(x,z) e^{-iq_x x - iq_z z} = -\frac{2i}{q_x} P_0 \hat{\boldsymbol{z}} \tag{3}$$

Hence,

$$\widetilde{\varphi}_{q_x, q_z} = -\frac{8\pi P_0}{\varepsilon} \frac{q_z}{q_x} \frac{1}{q_x^2 + q_z^2} \tag{4}$$

Evaluating the inverse Fourier transform we obtain

$$\varphi(x,z) = -\frac{2P_0}{\varepsilon} i \ln \frac{-x + iz}{x + iz} = \frac{2P_0}{\varepsilon} (\pi - 2\alpha) \tag{5}$$

where $\alpha$ is the polar angle of the vector $(x,z)$. The electric field components read

$$E_x = -\frac{4P_0}{\varepsilon} \frac{z}{x^2 + z^2}, \, E_y = 0, \, E_z = \frac{4P_0}{\varepsilon} \frac{x}{x^2 + z^2} \tag{6}$$